\documentclass[conference]{IEEEtran}
\usepackage{amsmath,amssymb,amsfonts}
\usepackage{algorithm}
\usepackage{algpseudocode}
\usepackage{graphicx}
\usepackage{orcidlink}
\usepackage[detect-all]{siunitx}
\usepackage{textcomp}
\usepackage{xcolor}
\usepackage[numbers,sort&compress]{natbib}
\def\BibTeX{{\rm B\kern-.05em{\sc i\kern-.025em b}\kern-.08em
    T\kern-.1667em\lower.7ex\hbox{E}\kern-.125emX}}

\begin{document}

\title{GymD2D: A Device-to-Device Underlay Cellular Offload Evaluation Platform}

\author{\IEEEauthorblockN{David Cotton,\orcidlink{0000-0002-8817-3736}}
\IEEEauthorblockA{\textit{School of Electrical and Data Engineering} \\
\textit{University of Technology Sydney}\\
Sydney, Australia \\
david.f.cotton@student.uts.edu.au}
\and
\IEEEauthorblockN{Zenon Chaczko}
\IEEEauthorblockA{\textit{School of Electrical and Data Engineering} \\
\textit{University of Technology Sydney}\\
Sydney, Australia \\
zenon.chaczko@uts.edu.au}
}

\IEEEoverridecommandlockouts

\IEEEpubid{\begin{minipage}{\textwidth}\ \\[12pt]
  \copyright2021 IEEE. Personal use of this material is permitted. Permission from \\
  IEEE must be obtained for all other uses, in any current or future media, \\
  including reprinting/republishing this material for advertising or promotional \\
  purposes, creating new collective works, for resale or redistribution to servers \\
  or lists, or reuse of any copyrighted component of this work in other works.
\end{minipage}}

\maketitle
\IEEEpubidadjcol

\begin{abstract}
    Cellular offloading in device-to-device communication is a challenging optimisation problem in which the improved allocation of radio resources can increase spectral efficiency, energy efficiency, throughout and reduce latency.
    The academic community have explored different optimisation methods on these problems and initial results are encouraging.
    However, there exists significant friction in the lack of a simple, configurable, open-source framework for cellular offload research.
    Prior research utilises a variety of network simulators and system models, making it difficult to compare results.
    In this paper we present GymD2D, a framework for experimentation with physical layer resource allocation problems in device-to-device communication.
    GymD2D allows users to simulate a variety of cellular offload scenarios and to extend its behaviour to meet their research needs.
    GymD2D provides researchers an evaluation platform to compare, share and build upon previous research.
    We evaluated GymD2D with state-of-the-art deep reinforcement learning and demonstrate these algorithms provide significant efficiency improvements.
\end{abstract}

\begin{IEEEkeywords}
device-to-device (D2D) communication, cellular offload, resource allocation, radio resource management, network simulator, deep reinforcement learning, OpenAI Gym
\end{IEEEkeywords}

\section{Introduction}\label{sec:introduction}
Wireless data use is increasing rapidly presenting a challenge for cellular service providers.
The explosion of video and smart device traffic has intensified demand for limited cellular resources~\citep{rappaport2014mobile}.
A multi-faceted approach to developing next generation networks include: unlocking millimeter-wave frequencies, increasing cell density, multiple-input multiple-output, improved information encoding, and the focus of this research---smarter coordination~\citep{gupta2015survey}.

Device-to-device (D2D) communication is a broad set of protocols for ad-hoc, peer-to-peer wireless communication for cellular or Internet connected wireless devices.
In contrast to normal cellular operation, user equipment (UE) utilising D2D mode communicate directly with one another instead of through base stations and connected networks~\citep{kaufman2008cellular}.
D2D has been proposed for a variety of use cases such as: public safety communications, communications relay, localised services, Internet of things and cellular offload~\citep{asadi2014survey}.

\begin{figure}[ht]
    \includegraphics[width=\columnwidth]{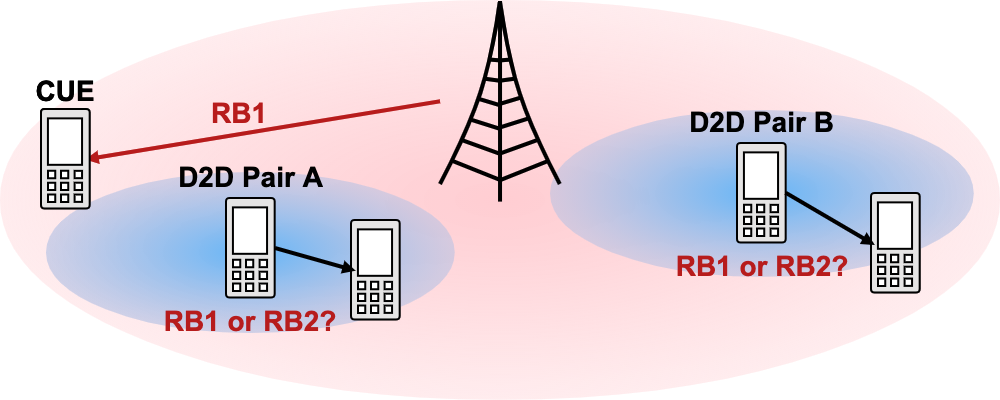}
    \caption{\textbf{Example of D2D cellular offloading with underlay networks.}
    This example demonstrates the cellular offload optimisation problem.
    Depicted is a BS, CUE and 2 DUE pairs.
    The BS has 2 RBs available to allocate.
    The BS has assigned the CUE RB1 for downlink and has the option of either assigning pair A RB1 and pair B RB2 or vice versa.
    If the BS assigns RB1 to pair A it is likely that due to their proximity to the CUE, significant interference would occur.
    As pair B is situated relatively distant to the CUE, if pair B transmitted with a lower power level, their interference on the CUE would be negligible.
    Therefore the optimal solution is to assign RB1 to pair B and RB2 to pair A.}
    \label{fig:cellular_offload}
\end{figure}

In this research we are interested in cellular offload, a mechanism to divert cellular traffic through alternative channels to improve network efficiency.
This can be achieved by communicating out of band, such as over WiFi, or inband via overlay or underlay networking.
Underlay cellular offload is a form of opportunistic spectrum access in which D2D UE (DUE) act as secondary users sharing radio resources with primary cellular UE (CUE)~\citep{janis2009interference}.
In underlay networking, DUE are responsible for managing their interference with CUE so as to avoid degrading primary network performance.

Radio resource management (RRM) is the system wide management of radio resources, such as transmit power and resource blocks (RB), across a wireless network to manage interference and utilise resources as efficiently as possible.
In cellular systems, which are limited by co-channel interference, improved resource allocation can increase spectral efficiency, energy efficiency, throughput and reduce latency.
We show a simplified example of the resource allocation problem in Figure~\ref{fig:cellular_offload}.

A challenge facing researchers developing D2D algorithms is the need for improved tooling, reliable measurements and established benchmarks.
An open-source evaluation platform allows researchers to compare results, share algorithms and build upon previous research.

In this paper we present \textit{GymD2D}, a network simulator and evaluation platform for RRM in D2D underlay cellular offload.
GymD2D provides convenient abstractions to aid researchers in quick prototyping of resource allocation algorithms.
The toolkit allows users to programmatically configure the environment to simulate a wide variety of scenarios.
GymD2D has been designed with extensibility as a core design principle, allowing users to override and extend its behaviour to meet their research needs.
It has been developed in the Python programming language to allow users to leverage its extensive ecosystem of scientific computing packages.
The open-source nature of GymD2D centralises development effort and avoids the redundant work of individual researchers creating their own simulators.
This puts more eyes on bug fixing, provides a more stable platform and increases confidence in reported empirical results.
GymD2D reduces entry barriers for junior researchers, helps researchers from other disciplines to cross-pollinate ideas easier and more generally increases participation.
Our software package is provided to the community under a MIT licence at https://github.com/davidcotton/gym-d2d.

\section{Background}\label{sec:background}
\subsection{Device-to-device communication}\label{subsec:device-to-device-communication}
In this section we provide an overview of the D2D RRM literature to situate the reader as to the requirements of the platform.
Firstly, we highlight the most common optimisation problems.
Secondly, we analyse key differences across simulators, paying special attention to simplifying assumptions frequently observed.
Thirdly, we survey the optimisation algorithms used for resource allocation.
Finally, we outline limitations of existing research, providing direction for the simulation requirements of future work.

\subsubsection{Optimisation Problems}
RRM is an optimisation problem where the objective is to utilise radio resources as efficiently as possible.
D2D RRM has been proposed for improving spectral efficiency, energy efficiency and quality of service.
In these uses cases, the objective can be to optimise data rates, throughput, capacity, signal to interference noise ratio (SINR), power consumption, energy efficiency or latency~\citep{asadi2014survey,chakraborty2020comprehensive}.
D2D systems can be centrally managed by the network operator, manage interference autonomously or use a hybrid control mode which aims to combine the benefits of both.
The choice of control mode limits the applicability of certain algorithms which may only be feasible in centrally managed paradigms.

\subsubsection{Simulation models}
D2D cellular offload typically investigates networks using orthogonal frequency division multiple access (OFDMA), communicating on licensed bands using underlay networking.
The most common scenario is a single macro base stations (MBS) surrounded by many randomly positioned CUEs and DUEs.
It is generally assumed that cellular systems are under full load and each RB is allocated to a CUE.
In the literature, simulations vary in scope from 2--30 RBs, 2--30 CUEs and 2--60 DUEs, while MBS operate with a cell radius of 20--500m.
It is frequently assumed that DUE are already paired and operate in a range between 10--30m apart.
Typically, omni-directional antenna and isotropic propagation are utilised.
Path loss is commonly modeled using log-distance models, with or without shadowing.

\subsubsection{Optimisation Algorithms}
A wide variety of optimisation methods have been investigated on a range of D2D RRM problems.
Initially, D2D radio resources were proposed to be managed using existing cellular uplink power control mechanisms~\citep{janis2009device}.
Consequently, it was identified that resources could be more efficiently allocated with the use of mathematical optimisation~\citep{yu2009power}.
However, due to the computational complexity of these methods and the millisecond timescales involved, it may not feasible to solve to optimality.
This can be addressed with the use of greedy heuristic algorithms which reduce computational complexity at the cost of global optimality~\citep{zulhasnine2010efficient}.
Alternatively, resource allocation can be optimised graph theoretically~\citep{zhang2013interference}, game theoretically~\citep{xu2012interference}, with evolutionary algorithms~\citep{su2013resource} or reinforcement learning (RL)~\citep{luo2014dynamic}.
More recently, deep reinforcement learning (DRL), a subfield of RL which uses deep neural networks to represent policy and/or value functions has demonstrated promising results~\citep{li2019multi,tan2019deep}.
DRL is well suited for many D2D RRM problems as neural networks provide rich approximations, scale well and generalise to unseen data.

\subsubsection{Research limitations}
A common limitation observed in D2D cellular offload research is BSs not enforcing uplink power control for CUE, who transmit at maximum power, a very energy inefficient approach.
Another research challenge is accounting for large SINR increases on the primary network, such as could significantly impact primary network throughput or drive up CUE transmit power levels.
Resource allocations algorithms need to demonstrate their effectiveness in larger search spaces that more closely reflect real world demands.
Iterative learning algorithms need to be capable of generalising to out of training distribution data and be robust under diverse propagation conditions.
Lastly, in our opinion, one of the greatest limitations of existing research is the lack of established benchmarks and comparison with other algorithms.

\subsection{OpenAI Gym}\label{subsec:openai-gym}
OpenAI Gym is an open-source software toolkit for reinforcement learning (RL)~\citep{brockman2016openai}.
Gym provides an abstraction layer that enables a variety of tasks, known as \textit{environments} in RL parlance, to be wrapped to present a consistent interface.
The abstraction provided by Gym allows the easy interchange of algorithms and environments.
This makes it is easy to test how a single algorithm generalises across a diverse set of environments or to benchmark different algorithms on a given environment.
The simplicity and flexibility Gym offers has proved very popular and has lead to it becoming the de facto environment format in RL.
While Gym was designed for RL research, the application programming interface (API) it provides makes it easy to apply many other algorithms types.

\subsection{Network simulation}\label{subsec:network-simulation}
%

One of the most widely used network simulators in education and research is \textit{ns-3}.
Ns-3 is an open-source, modular, discrete-event simulator for wired and wireless networks.
It provides the full TCP/IP stack and wireless propagation modelling.
Another popular alternative with comparable features is \textit{OMNeT++}, while there exists similar commercial tools such as \textit{NetSim} and \textit{MATLAB}.
Ns-3 has been incorporated into an OpenAI Gym environment under the \textit{ns3-gym} project~\citep{gawlowicz2018ns3}.

\section{Design Principles}\label{sec:design-principles}
The design of GymD2D has been inspired by the authors experience developing and comparing reinforcement learning algorithms.
In our experience the following design principles stimulate experimentation and the sharing of ideas.

\begin{itemize}
    \item \textbf{Simple}: Easy to get started with, the framework should allow researchers to be productive quickly.
    \item \textbf{Configurable}: The framework should be easily configured to meet the broad range of D2D cellular offload use cases.
    Configurability allows researchers to programmatically test algorithm generalisation and scalability.
    \item \textbf{Extensible}: The framework should allow users to extend the system's behaviour to meet their needs.
    The nature of research dictates a stream of new ideas we can't anticipate but we can provide researchers the flexibility to adapt.
    \item \textbf{Scalable}: The framework should be performant and easily parallelisable.
    Developing new algorithms requires significant experimentation and reducing the time spent waiting for results is important for productivity.
    Some algorithms, such as policy gradient DRL, require parallel environments to function.
    Real world solutions are often a combination of both algorithmic and architectural components.
    \item \textbf{Reproducible}: Experiments should be easily repeatable.
    To build confidence in our deductions, it is important that we can reperform experiments to ensure the observed outcomes were not statistical anomalies.
    Reproducibility allows researchers to share their contributions with community more easily.
\end{itemize}

\section{System Design}\label{sec:system-design}

\subsection{System model}
GymD2D is designed to simulate physical layer resource allocation problems in D2D underlay cellular offload.
The framework abstracts away data link and above layers, D2D session establishment and management concerns.
The system model contains a single MBS $b$, a set of $M$ CUEs $\mathcal{M} = \{1,\dots,M\}$ and a set of $N$ DUE pairs $\mathcal{N} = \{1,\dots,N\}$, that reside within the coverage area of the cell.
We denote the $m^{th}$ CUE $C_{m}$, the $n^{th}$ DUE pair $D_{n}$ and the transmitter and receiver of pair $D_{n}$ by $D_{n}^{t}$ and $D_{n}^{r}$ respectively.

The system employs OFDMA, with a set of $K$ RBs $k \in \mathcal{K}$ are available for allocation.
An assumption is made that all devices are equipped with omni-directional antenna and transmit isotropically.
Accordingly, the network resides within a circular cell of radius $R$, with the MBS located in the centre at position $(0,0)$.
The simulation environment contains no obstructions or outside interference.
D2D communicate one-to-one and D2D relay is not supported.

We denote the effective isotropic radiated power (EIRP) $P$ of BSs, CUEs and DUEs as $P^{b}$, $P^{c}$, $P^{d}$ respectively.
The EIRP of a BS is calculated,
\begin{equation}
    P^{b} = P_{tx} - 10log_{10}s + g_{ant} - l_{ix} - l_{cb} + g_{amp}
\end{equation}
and the EIRP of CUE and DUE,
\begin{equation}
    P^{c} = P^{d} = P_{tx} - 10log_{10}s + g_{ant} - l_{ix} - l_{bd}
\end{equation}
where $P_{tx}$ is the transmission power level, $s$ is the number of subcarriers, $g_{ant}$ is the transmitting antenna gain, $l_{ix}$ is the interference margin loss to approximate noise from surrounding cells, $l_{bd}$ is body loss to approximate attenuation caused by the user, $l_{cb}$ is cable loss, and $g_{amp}$ is amplifier gain.

We denote the received signal level $R$ from transmitter $i$ at receiver $j$ of BS, CUEs and DUEs as $R^{b}_{i,j}$, $R^{c}_{i,j}$, $R^{d}_{i,j}$.
The received signal level of BS as,
\begin{equation}
    R^{b}_{i,j} = P_{i} - PL_{i,j} + g_{ant} - l_{cb} + g_{amp}
\end{equation}
and the received signal level of CUE or DUE,
\begin{equation}
    R^{c}_{i,j} = R^{d}_{i,j} = P_{i} - PL_{i,j} + g_{ant} - l_{bd}
\end{equation}
where $P_{i}$ is the EIRP from transmitter $i$ and $PL_{i,j}$ is the path loss of the chosen path loss model between $i$ and $j$.

We assume D2D transmissions are synchronised to cellular transmissions and occupy the same $K$ orthogonal resources.
During both uplink and downlink, co-channel interference is calculated for each receiver sharing RB $k$.
GymD2D considers co-channel interference between:
\begin{itemize}
    \item \textit{D2D to cellular}, interference from secondary DUE on the primary cellular network,
    \item \textit{cellular to D2D}, interference from CUE or BS to DUE, and
    \item \textit{D2D to D2D}, the interference between DUE pairs sharing a RB.
\end{itemize}

Accordingly, we model the instantaneous SINR $\xi$ of receiver $j$ from transmitter $i$ on RB $k$,
\begin{equation}
    \xi_{i,j,k} = \frac{R_{i,j}}{\sum_{n \in \mathcal{T}_{k}, n \neq i} R_{n,j} + \sigma^{2}}
\end{equation}
where $\mathcal{T}_{k}$ is the set of transmitters allocated to RB $k$ and $\sigma^{2}$ is additive white Gaussian noise (AWGN).

The capacity of channel $C_{i,j}$ can be calculated using the SINR $\xi_{i,j}$,
\begin{equation}
    C_{i,j}[Mbps] = B log_{2}(1 + \xi_{i,j})\label{eq:shannon}
\end{equation}
where $B$ is the channel bandwidth in MHz.

\subsection{Path loss models}\label{subsec:path-loss}
GymD2D contains several of the most common path loss models and makes it easy for users to implement their own custom models.
By default, GymD2D uses the simplest model, free space path loss (FSPL),
\begin{equation}
    FSPL(f,d)[dB] = 10nlog_{10}\Big( \frac{4 \pi fd}{c} \bigg)\label{eq:ldpl}
\end{equation}
where $n=2$ is the path loss exponent (PLE) in free space, $f$ is the carrier frequency in Hz, $d$ is the distance between the transmitter and receiver and $c$ is the speed of light in m/s.

To simulate obstructed propagation environments it can be useful to model fading effects as random processes.
One such model is the log-distance with shadowing path loss model, which is included in GymD2D.
The log-distance path loss model extends FSPL to mimic random shadowing effects, such as caused by buildings, with a log-normal distribution,
\begin{equation}
    PL^{LD}(f,d)[dB] = FSPL(f, d_{0}) + 10nlog_{10}\frac{d}{d_{0}} + \chi_{\sigma}\label{eq:ldpl-shadowing}
\end{equation}
where $d_{0}$ is an arbitrary close-in reference distance, typically 1--100m and $\chi_{\sigma}$ is a zero-mean Gaussian with standard deviation $\sigma$ in dB.
Empirical measurements have shown values of $n = 2.7 \text{ to } 3.5$ to be suitable to model urban environments~\citep{rappaport1996wireless}.


\subsection{Architecture}\label{subsec:architecture}
\begin{figure}[ht]
    \includegraphics[width=\columnwidth]{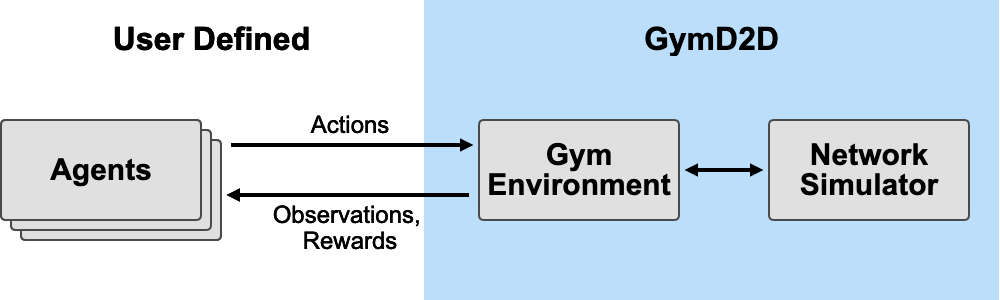}
    \caption{\textbf{Proposed GymD2D architecture.}
    GymD2D consists of a network simulator, wrapped by an OpenAI Gym environment.
    The user creates their own RRM algorithms to control wireless devices.}
    \label{fig:architecture}
\end{figure}

GymD2D consists of two main components, a network simulator and a Gym environment.
The network simulator models physical layer cellular networking.
The Gym environment provides an abstraction layer to allow researchers to experiment with different simulation parameters and algorithms programmatically.
Users supply RRM algorithms to manage the wireless devices under simulation.
GymD2D outputs data on the state of the simulation to the user;
to allow the effectiveness of RRM algorithms to be studied, such as through visualisation.
A high level overview of the architecture of GymD2D is depicted in Figure~\ref{fig:architecture}.

\subsection{Network simulator}\label{subsec:network-simulator}
The network simulator models a single cellular cell which is populated with a collection of randomly placed CUEs and DUE pairs.
It is a configurable component which can be customised to emulate a range of cellular offload scenarios.
This includes the number and configuration of BSs, CUEs and DUEs and environmental parameters such as the available RBs, cell size and path loss model.

\begin{figure}[ht]
    \includegraphics[width=\columnwidth]{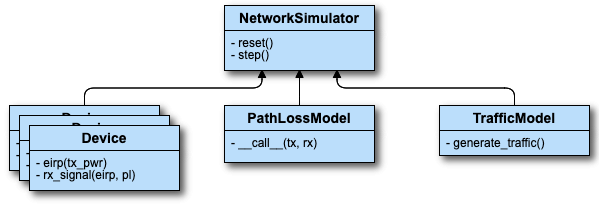}
    \caption{\textbf{Network simulator architecture.}
    The main components of the network simulator are a collection of CUEs, DUEs and BS, the path loss model and the traffic model.}
    \label{fig:simulator_cls}
\end{figure}

The main components of the network simulator are: a collection of wireless devices (BSs, CUEs, DUEs), a path loss model and a traffic model as shown in the class diagram in Figure~\ref{fig:simulator_cls}.

Each simulation, the actions of BS and UEs within the cell can be generated internally by the \textit{traffic model} or externally from a user defined RRM algorithm.
A typical use case would be to use the internal traffic model to control BS and CUEs and the user RRM algorithm the DUEs.

GymD2D uses a discrete-event simulation model.
This method is congruent with the Gym API in which the incoming actions are the \textit{events} and the Gym \textit{step()} method calls equate to the system update intervals and model a single LTE or NR frame.
At each step, each device may transmit, receive, or take no action.
An action is tuple consisting of a transmitter, receiver, communication mode, RB and transmission power.
The simulator consolidates the actions from both the traffic model and the RRM algorithm, then calculates the resulting propagation and interference.
After calculating propagation, metrics on the state of the network, such as SINR and throughput, are output to the Gym environment.

\subsection{Gym environment}\label{subsec:gym-environment}
The Gym environment has been designed to be configuration driven, to facilitate the programmatic scheduling and reproducibility of experiments.
When instantiating a new Gym environment, configuration can be provided to specify the BSs, CUEs and DUEs that inhabit the simulation and the environmental conditions.

The Gym environment provides RRM algorithms with an observation and action space.
These provide a mapping to configure for the expected format of inputs and outputs.
For example, when using a DRL algorithm, this would allow DRL to configure its neural networks for the shape of incoming observations and output actions of the correct dimension.

At each step, the Gym environment receives actions from the RRM algorithm and converts them to a format suitable for the network simulator.
Once a simulation step is complete, the environment uses the state of the simulator to create the observations and rewards RRM algorithms consume to make their decisions.

\section{Evaluation}\label{sec:evaluation}
\subsection{Methodology}\label{subsec:methodology}
\begin{figure*}
\includegraphics[width=\textwidth]{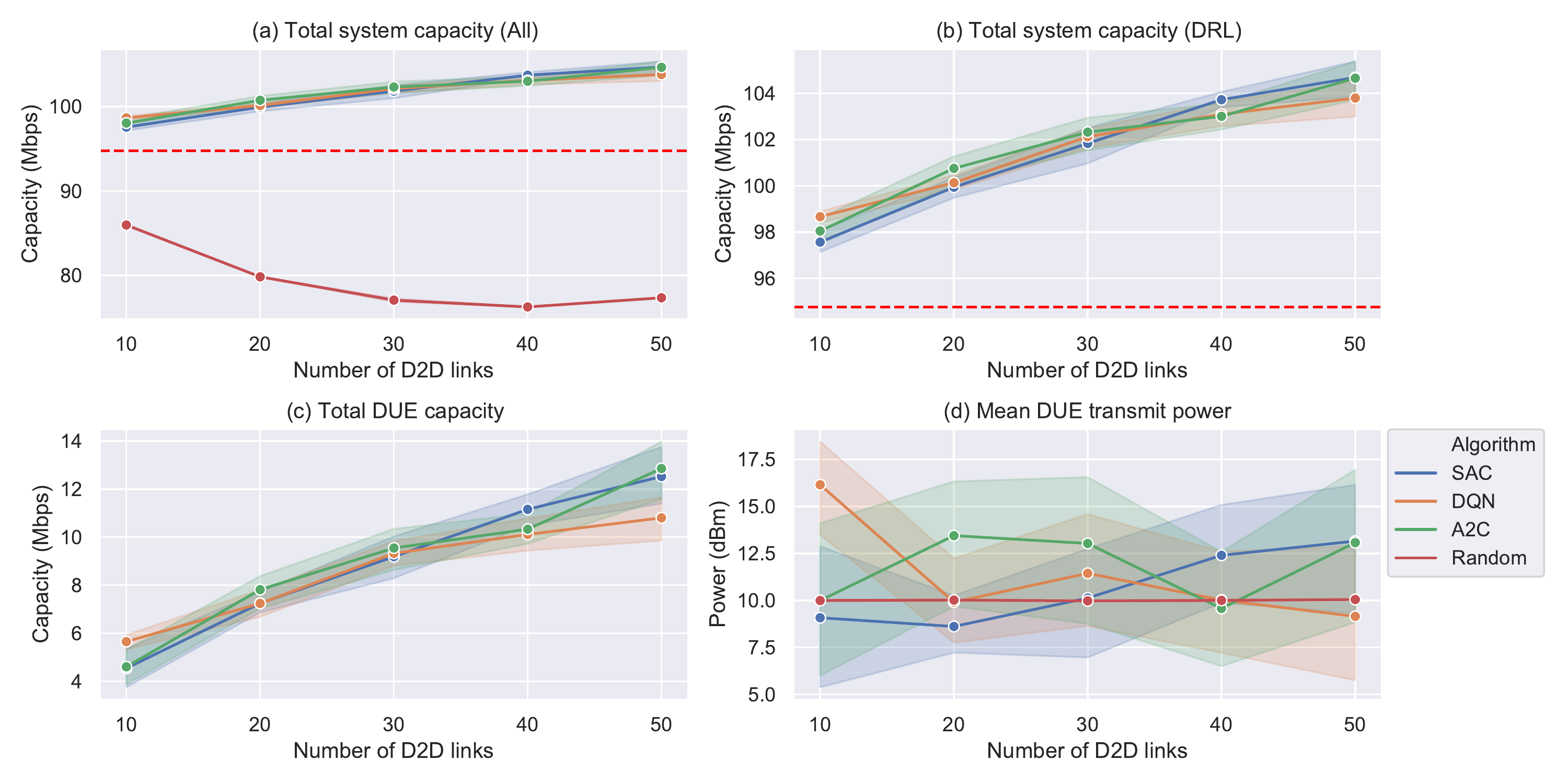}
\caption{\textbf{Evaluation results.} To evaluate GymD2D we compared the performance of several state-of-the-art DRL algorithms in their efficiency allocating radio resources as D2D demand increased. Solid lines indicate the mean algorithm performance across ten trials with the shaded area the 95\% confidence interval. The dashed red line indicates the baseline total system capacity without D2D communication. (a) The total system capacity of the DRL and random agent. (b) The total system capacity of just the DRL agents. (c) The total DUE capacity of the DRL agents. (d) The mean transmit power of all agents.}
\label{fig:evaluation}
\end{figure*}

We evaluated GymD2D with several leading DRL algorithms to determine their efficiency allocating radio resources as D2D demand increased.
The objective was to maximise the total system capacity, that is the sum data rate of all CUE and DUE, calculated for each transmitter/receiver pair $i,j$ by
\begin{equation}
    C_{i,j}[Mbps] =
    \begin{cases}
        B log_{2}(1 + \xi_{i,j}) &  \xi_{i,j} \geq \rho_{j} \\
        0 & \xi_{i,j} < \rho_{j} \\
    \end{cases}
\end{equation}
where $B=0.18$ is the RB bandwidth in MHz and $\rho_{b}=-123.4$ and $\rho_{d}=-107.5$ is the receiver sensitivity of a BS and DUE respectively in dBm.
Our evaluation simulated a single cell under full load.
The scenario contained 25 RBs and CUEs, with each CUE allocated an individual RB.
We employed a centrally managed control mode in which DUE communicated in the uplink frame, with the resource allocation managed by the network operator.
Each RRM algorithm was evaluated with 10, 20, 30, 40 and 50 communicating D2D pairs.
Algorithms were compared by training to convergence, then evaluating for 100 episodes.
For each algorithm--D2D link density comparison, we conducted ten trials, retraining from scratch and evaluating, to account for variations in performance.
Each episode lasted for ten steps or equivalently ten LTE/NR frames to simulate short bursts of traffic on a busy network.
In each episode all CUE and DUE remained geographically fixed, but at the end of each episode, all CUE and DUE were randomly repositioned within the cell to simulate new devices accessing the network.
Wireless propagation was modelled using the Log-Distance Shadowing model~\eqref{eq:ldpl-shadowing} with PLE $n=2.0$ and $\chi_{\sigma}=2.7$.
The simulation parameters are detailed in Table~\ref{tab:simulation_params}.

\begin{table}
\centering
\caption{Simulation parameters}
\label{tab:simulation_params}
\begin{tabular}{c|c}
    \hline
    Parameter & Value \\
    \hline
    Cell radius & $500$ m \\
    Maximum D2D pair distance & $30$ m \\
    Carrier frequency & $2.1$ GHz \\
    RB bandwidth & $180$ kHz \\
    Number of RBs & $25$ \\
    Number of CUEs & $25$ \\
    Number of DUE pairs & $10, 20, 30, 40, 50$ \\
    CUE transmit power & $23$ dBm \\
    DUE min, max transmit power & $0, 20$ dBm \\
    Path loss model & Log-Distance Shadowing \\
    Path loss exponent & $2.0$ \\
    Shadowing SD $\chi_{\sigma}$ & 2.7 \\
\end{tabular}
\end{table}

We evaluated three DRL algorithms, Rainbow DQN~\citep{hessel2017rainbow}, Discrete Soft Actor-Critic (SAC)~\citep{christodoulou2019soft}, and Advantage Actor-Critic (A2C)~\citep{mnih2016asynchronous}, and a random agent baseline.
All DRL algorithms used a fully connected neural network with two hidden layers trained using the Adam optimiser.
Each hidden layer contained 128 units and used ReLU activation between layers.
Policies used a reward discounting factor of $\gamma=0.9$.
Our Rainbow DQN (Table~\ref{tab:rainbow_hyperparams}) implementation used distributional, dueling, double-Q and noisy networks with a prioritised replay buffer and single step returns.
The discrete action space variant of the SAC (Table~\ref{tab:sac_hyperparams}) was used.
A2C (Table~\ref{tab:a2c_hyperparams}) is the synchronous version of A3C and used Generalised Advantage Estimator (GAE)~\citep{schulman2015high} with $\lambda=1.0$.

\begin{table}
\centering
\caption{Rainbow DQN hyperparameters}
\label{tab:rainbow_hyperparams}
\begin{tabular}{c|c}
    \hline
    Parameter & Value \\
    \hline
    Discounting factor $\gamma$ & $0.9$ \\
    Learning rate $\alpha$ & $5\cdot10^{-4}$ \\
    Batch size & $32$ \\
    Online network update period & $4$ steps \\
    Learning start & $1,000$ steps \\
    Target network sync period & $500$ steps \\
    Distributional atoms & 51 \\
    Distributional bounds $v_{min}$, $v_{max}$ & [-1,10] \\
    Replay buffer capacity & $50,000$ \\
    Replay buffer prioritisation exponent $\omega$ & $0.6$ \\
    Replay buffer importance sampling $\beta$ & $0.6 \rightarrow 0.4$ \\
    Importance sampling annealing & 20,000 steps \\
\end{tabular}
\end{table}

\begin{table}
\centering
\caption{SAC hyperparameters}
\label{tab:sac_hyperparams}
\begin{tabular}{c|c}
    \hline
    Parameter & Value \\
    \hline
    Discounting factor $\gamma$ & $0.9$ \\
    Learning rate $\alpha$ & $3\cdot10^{-4}$ \\
    Batch size & $256$ \\
    Learning start & $1,500$ steps \\
    Target smoothing coefficient & 0.005 \\
    Replay buffer capacity & $50,000$ \\
\end{tabular}
\end{table}

\begin{table}
\centering
\caption{A2C hyperparameters}
\label{tab:a2c_hyperparams}
\begin{tabular}{c|c}
    \hline
    Parameter & Value \\
    \hline
    Discounting factor $\gamma$ & $0.9$ \\
    Learning rate $\alpha$ & $1\cdot10^{-4}$ \\
    Rollout length & $10$ \\
    Entropy coefficient $\beta$ & $0.01$ \\
    GAE $\lambda$ & $1.0$ \\
\end{tabular}
\end{table}

\subsection{Results}\label{subsec:results}
The results of our evaluation can be seen in Figure~\ref{fig:evaluation}.
Baseline total system capacity measures the efficiency of the system without D2D communication.
For our scenario this was 94.75 Mbps.
We found that all the DRL algorithms achieved a similar level of performance, increasing system capacity over the baseline by more than 11\%.
Our results show that the system capacity continued to increase sublinearly as number of active D2D links demand grew.
Conversely, the performance of the random agent shows that without careful resource allocation, the system capacity drops sharply.

We found that despite allowing DUE to communicate up to half the power of CUE (20 vs.\ 23 dBm), they typically converged into operating ranges between 7 and 15 dBm.
This resulted in a negligible decrease in the total CUE capacity, 1--2 Mbps or $\approx$1.84\% below the baseline system capacity.
This decrease was approximately constant across D2D density.


\subsection{Discussion}\label{subsec:discussion}

Inspecting the actions the DRL algorithms selected, we found that they converged to allocating all DUE onto one or two RBs.
This is surprising as we had anticipated the DUE to be evenly distributed amongst all available RBs.
Investigating further, we observed that over the course of a training run, the DQN converging from an even RB distribution to the focused allocation strategy.
This behaviour developed in the later stages of training and only contributed modest increases to the system capacity.

As expected, the optimal strategy for resource allocation was to assign DUE to share RBs with the most geographically distant CUE\@.
When combined with the focused RB allocation described above, this typically resulted in the RRM algorithm choosing to allocate DUE to share with the one or two most isolated CUE.

Despite the random UE positioning, the DRL agents were able to learn policies that generalised much better than we anticipated when using fully connected neural networks.
We were also surprised how quickly agents adapted during an episode, improving their performance over the course of the ten-step episode.




\section{Conclusion}\label{sec:conclusion}
In this research we have presented GymD2D, a network simulator and evaluation platform for RRM in D2D underlay cellular offload.
GymD2D makes it easy for researchers to build, benchmark and share RRM algorithms and results.
Our toolkit is designed to quickly prototype physical layer resource allocation algorithms, without the complexity of higher layer protocols.
GymD2D is configurable and extensible, allowing it to be employed to simulate a range of D2D research needs.

We have evaluated GymD2D with several leading DRL algorithms and demonstrated the performance gains of intelligent RRM, increasing system capacity by more than 11\%.
There was no clear winner amongst the DRL algorithms which performed similarly.
The results also demonstrated that D2D cellular offload can significantly minimise its impact on primary networks.

In the future we plan to increase the simulation complexity in GymD2D, adding more realistic modelling.
Other interesting research challenge include investigating the impacts of CUE power control on cellular offload and supporting D2D relay.
We continue to use GymD2D in ongoing research, developing methods for scaling up DRL based D2D RRM.

\bibliographystyle{IEEEtran}
\bibliography{bibfile}

\end{document}